\documentclass[runningheads]{svmult}

\usepackage{makeidx}   % allows index generation
\usepackage{graphicx}  % standard LaTeX graphics tool
\usepackage{subeqnar}  % subnumbers individual equations
\usepackage{multicol}  % used for the two-column index
\usepackage{physprbb}  % modified textarea for proceedings,
\makeindex             % used for the subject index
\def\msun{\ifmmode M_{\odot} \else M$_{\odot}$\fi}
\def\zsun{\ifmmode Z_{\odot} \else Z$_{\odot}$\fi}
\def\lsun{\ifmmode L_{\odot} \else L$_{\odot}$\fi}
\def\teff{\ifmmode T_{\rm eff} \else $T_{\rm eff}$\fi}
\def\mup{$M_{\rm{up}}$}
\def\micron{$\mu$m}

\def\etal{et al.}
\def\hii{H~{\sc ii}}
\def\lya{Ly$\alpha$}

\def\la{\raisebox{-0.5ex}{$\,\stackrel{<}{\scriptstyle\sim}\,$}}
\def\ga{\raisebox{-0.5ex}{$\,\stackrel{>}{\scriptstyle\sim}\,$}}
\def\ii{\'{\i}}

\newcommand{\atom}[2]{#1~{\sc #2}}

\def\aap{A\&A}
\def\aaps{A\&AS}
\def\apj{ApJ}
\def\apjs{ApJS}
\def\mnras{MNRAS}
\begin{document}
\title*{Stellar features in integrated starburst spectra
as stellar population diagnostics\thanks{Invited review to 
appear in ``Starbursts -- near and far'', Eds. L. Tacconi-Garman,
D. Lutz, Springer Verlag}}
%
%
%\toctitle{Focusing of a Parallel Beam to Form a Point
%\protect\newline in the Particle Deflection Plane}
% allows explicit linebreak for the table of content
%
%
%\titlerunning{Focusing of a Parallel Beam}
% allows abbreviation of title, if the full title is too long
% to fit in the running head
%
\author{Daniel Schaerer\inst{1}}
%
%\authorrunning{Ivar Ekeland et al.}
% if there are more than two authors,
% please abbreviate author list for running head
%
%
\institute{Laboratoire d'Astrophysique, Observatoire Midi-Pyr\'en\'ees, 
       14, Av. E. Belin, F-31400, Toulouse, France
      (schaerer@ast.obs-mip.fr)}
\maketitle              % typesets the title of the contribution

\begin{abstract}
We review the main stellar features observed in starburst spectra from 
the UV to the near-IR and their use as fundamental tools to 
determine the properties of stellar populations from integrated spectra.
The origin and dependence of the features on stellar properties are discussed, 
and we summarise existing modeling techniques used for quantitative
analysis. 
Recent results from studies based on UV, optical and near-IR observations
of starbursts and active galaxies are summarised.
Finally, we briefly discuss combined starburst + photoionisation models
including also observations from nebular emission lines.
The present review is complementary to the recent summary by  
Schaerer (2000) discussing more extensively nebular analysis of
starbursts and related objects.

\end{abstract}
%%%%%%%%%%%%%%%%%%%%%%%%%%%%%%%%%%%%%%%%%%%%%%%%%%%%%%%%%%%%%%%
\section{Introduction}
The analysis of distinct spectral features in integrated spectra 
is at the base of numerous investigations on the stellar content 
of distant galaxies.
Indeed, in addition to the overall continuum spectral shape, stellar
absorption (and rarely also emission) lines carry crucial 
information on the presence of stars of various spectral types and
luminosity class, and allow thus in principle to ``decompose''
the integrated galaxy spectrum in its stellar constituents, 
and to determine their fundamental properties such as ages,
IMF, the star formation history etc.

In objects such as active galaxies, where non stellar emission processes
are thought to contribute to the emitted light, the study of possible 
stellar features allows one to constrain the relative stellar contribution,
and thus to determine the efficiency of various emission processes
(e.g.\ stellar versus non-stellar).

These basic properties illustrate the interest of spectroscopic
studies of stellar features in starbursts and other galaxies.

The aim of the present review is to discuss the main stellar features
observed in starburst spectra over the entire spectral range where such
features are detectable, i.e.\ from the UV over the optical to the 
near-IR. At longer wavelength dust emission dominates and the stellar 
continuum and associated lines is not detectable anymore.

An ``inventory'' of the strongest stellar features is given for each 
spectral domain and recent results in the respective fields are summarised.
In the last Section, we also briefly discuss the use of combined
stellar and nebular emission line (hereafter EL) analysis for
the study of stellar populations in the otical and IR.

The current review is complementary to a recent review 
Schaerer (2000) discussing
new developments in multi-wavelength modeling tools, the current
status of ionising fluxes from massive stars, and their importance
for EL analysis of starbursts and related ojects.

%%%%%%%%%%%%%%%%%%%%%%%%%%%%%%%%%%%%%%%%%%%%%%%%%%%%%%%%%%%%%%%
\section{UV features}
The UV spectral range ($\sim$ 1000 -- 3000 \AA) of starbursts is 
rich in stellar lines originating in early type stars (mostly OB, 
also Wolf-Rayet), it contains few or weak nebular lines, and
rather numerous interstellar (IS) absorption lines.

A rough inventory of the strongest {\em stellar} lines observed 
mostly in the $\sim$ 1200 -- 1800 \AA\ range follows
(cf.\ the detailed work of de Mello \etal\ 2000, hereafter DLH00):
% for identifications and discussion for the 1200 -- 1800 \AA\
%range):
\begin{itemize}
\item Well known stellar wind lines (P-Cygni or EL) from O and Wolf-Rayet (WR) stars)
are  \atom{Si}{iv} 1400, \atom{C}{iv} 1550, \atom{N}{v} 1240, \atom{He}{ii} 1640, 
and \atom{N}{iv}1720. The following synthesis models include at least partly these 
lines and discuss their behaviour: 
Sekiguchi \& Anderson (1987), Mas-Hesse \& Kunth (1991),
Fanelli \etal\ (e.g.\ 1987, 1992), Leitherer, Robert \etal\ (1993-2000).

\item Other wind lines blueward of \lya\ include \atom{O}{vi}+Ly$\beta$+\atom{C}{ii} 
1010-1060, discussed and modeled by Gonz\'alez Delgado \etal\ (1999), and potentially
lines of C, N, P, S, and Ar in the range recently observed with FUSE
(see Taresch \etal\ 1997, Fullerton \etal\ 2000).

\item The strongest photospheric lines from OB stars are \atom{Si}{ii} 1265, 1485, 
\atom{Si}{iii}1295-1300, 1417, \atom{C}{ii} 1334, 1335, \atom{C}{iii} 1247, 1427, 
\atom{S}{v} 1501 (see de Mello \etal\ 2000).

\item Other stellar features include ``depressions'' due to numerous Fe lines 
  (at $\sim$ 1400, 1600, 1940), and also \atom{Fe}{ii} 2570-2615 and 
  \atom{Mg}{ii} 2780-2825 features at longer wavelengths 
  (e.g.\ Robert \etal\ 1999, Storchi-Bergmann \etal\ 1995)
\end{itemize}
It is important to note that many of these lines can also be formed in the
interstellar medium (cf.\ Heckman \& Leitherer 1997, Sahu 1998). 
A careful separation of the stellar and interstellar component is necessary
in many cases (see de Mello \etal\ 2000).
The wavelength range between $\sim$ 1800 and 3000 \AA\ remains still
little explored. A similarly detailed understanding of this spectral range
is of importance for studies of galaxies in the $z \sim$ 1--2 redshift
range (see e.g.\ Campusano \etal\ 2000).

Given the dependence of the various lines on stellar luminosity, age, and also metallicity
(cf.\ below), e.g.\ well studied for the stellar wind lines (e.g.\ Walborn \etal\ 1985, 
Leitherer \& Lamers 1991), the features can be used to constrain the parameters
of the integrated  population, such as age, SF history, and IMF, by means of evolutionary
synthesis techniques (e.g.\ Mas-Hesse \& Kunth 1991, Leitherer \etal\ 1995). 
The most up-to-date model suited to such analysis is {\em Starburst99} (Leitherer \etal\
1999, de Mello \etal\ 2000).

These techniques have been extensively applied to the interpretation of
UV spectra of nearby starbursts (mostly HST spectra), especially by Leitherer,
Heckman, Gonz\'alez Delgado and collaborators (some references given below)
and by Mas-Hesse \& Kunth (1991, 1999). 
%To summarise the main result of these
%studies in one sentence the main result of these studies 
%
Summarised in one sentence ($\ldots$) the main result of these studies is that
all the objects contain young bursts (\la\ 10--20 Myr) characterised by 
instantaneous burst or continuous star formation, the distinction being often
difficult to draw, which are populated with a rather normal Salpeter-like IMF
with stars up to  \mup\ $\sim$ 60 -- 100 \msun.
In a recent study Tremonti \etal\ (2000, cf.\ these proceedings) examine
the stellar populations in the field of NGC 5253 and find a possible indication
for a steeper IMF, although other explanations (e.g.\ age effects) are possible.

The similarity of the spectra of many high redshift galaxies (e.g.\ Lyman break galaxies)
with the local starbursts is now well recognised and offers many exciting possibilities.
For example, from the beautiful spectrum of the lensed z $\sim$ 2.7 Galaxy 1512-cb58
of Pettini \etal\ (2000) these authors and de Mello \etal\ (2000) derive 
a constant star formation, an IMF slope between Salpeter (2.35) and $\sim$ 2.8,
and find indications for a subsolar metallicity, in agreement with
EL measurements from Tepliz \etal\ (2000).

Obviously it is of great interest to derive/estimate the metallicity ($Z$) from
stellar UV lines. Since, for example, the strength of stellar wind lines depend 
on $Z$ this is in principle possible. 
This can e.g.\ be done using the correlation of the equivalent width of \atom{Si}{iv}
with metallicity found by Heckman \etal\ (1998).
However, a priori, such a correlation should only be valid in a statistical sense,
since the line strength also depends on age (cf.\ Leitherer \etal\ 1995).
This difficulty should be less if the full wind line profile can be analysed.
The inclusion of spectral libraries of metal-poor stars in evolutionary synthesis
models has just been completed (e.g.\ Leitherer \etal\ 2000, see also Heap 2000).
Alternative possibilities to derive the chemical composition include the use
of IS absorption lines (cf.\ Pettini \etal\ 2000), or the use of weak stellar features
such as various Fe blends known to vary with $Z$ (cf.\ Haser \etal\ 1998).

%%%%%%%%%%%%%%%%%%%%%%%%%%%%%%%%%%%%%%%%%%%%%%%%%%%%%%%%%%%%%%%
\section{Stellar features in the optical}

The inventory of the strongest stellar lines in the optical is as follows:
\begin{itemize}
\item Broad emission lines from Wolf-Rayet stars of various subtypes (WN, WC) 
  are detected in some young starbursts: \atom{He}{ii} 4686
  bump, \atom{C}{iv} 5808,  \atom{C}{iii} 5696, possibly also \atom{N}{iii} 4512, 
  \atom{Si}{iii} 4565 (see e.g.\ Schaerer \etal\ 1999b, Guseva \etal\ 2000).
  Synthesis models treating these lines include
  Cervi\~no, \& Mas-Hesse (1994) and Schaerer \& Vacca (1998).
\item H and He absorption lines from OBA stars have been discussed and modeled
 by Diaz (1988), Olofsson (1995), and Gonz\'alez Delgado \etal\ (1999).
\item The \atom{Ca}{ii} triplet $\sim$ 8498, 8542, 8662 has often been studied
   (e.g.\ Terlevich \etal\ 1990ab, Mayya 1997, Garcia-Vargas \etal\ 1998).
  Its origin in both late type giants and supergiants complicates a priori 
   the analysis.
\item Other metallic features and molecular bands originating in stars with
  F types and later are
  \atom{Ca}{ii} H+K 39XX, CH G band 4284-4318, \atom{Mg}{i}+MgH 5156-5196, 
  \atom{Na}{i} 5880-5914, various TiO bands $\ge$ 6200.
  These are found e.g.\ in the template spectra of Bica \& Alloin (1986 and subsequent
  papers) for clusters and in the starburst spectra of Storchi-Bergmann \etal\ (1995).
\end{itemize}

A complete overview of all starburst studies exploiting these features
is not possible here. I shall instead briefly summarise recent results
on starburst (and possibly also AGN) studies using Wolf-Rayet features and 
metallic lines. The contribution of Gonz\'alez Delgado (these proceedings)
illustrates the use of H and He absorption lines.

\subsection{WR features as a probe of the most massive stars}
Since WR are the descendents of the most massive stars, detections of their 
features provide the best indication of the presence of massive stars
($M{\rm initial} \ga$ 25--60 \msun) and allow to constrain the upper
end of the IMF. A catalogue of all known galaxies with WR detections
has been compiled by Schaerer \etal\ (1999b; on the Web at
{\tt http://webast.ast.obs-mip.fr/people/schaerer/}).

\subsubsection{Starburst--AGN connection}
The detection of stellar features including the so-called WR-bump, UV lines and
the  \atom{Ca}{ii} triplet in the Seyfert2 galaxy Mrk 477 by Heckman \etal\ (1997)
has considerably re-vived this subject, before more focused on detections
of the \atom{Ca}{ii} triplet (e.g.\ Terlevich \etal\ 1990ab).
Since then other possible
WR detections indicating important massive star populations have been made
in Mrk 1210 (Sey2, Storchi-Bergmann \etal\ 1998), Mrk 463E and Mrk 1 (Sey2, Gonzalez 
Delgado \etal\ 2000), TF 1736+1122 (Sey 2, Tran \etal\ 1999), and in three PG QSO
(Lipari \etal\ 2000). Given the strong nebular contamination due to 
\atom{He}{ii} $\lambda$4686, other unambiguous massive star features are required 
to fully clearly establish the presence of massive stars in these objects.
Such attempts, aiming to detect the WR lines of \atom{C}{iv} and/or \atom{C}{iii} 
which are not affected by nebular contamination, have been undertaken by
Kunth \& Contini (1999) with 2-D spectroscopy in several Sey2.

From the analysis of WR and H+He absorption lines in their sample of 20 Sey2 
galaxies, Gonz\'alez Delgado \etal\ (2000; cf.\ these proceedings) find that 
the blue and near-UV light 
of half of their objects is dominated by young and/or intermediate age stars.
A similar result was found by Storchi-Bergmann \etal\ (1998) on a smaller sample.

\subsubsection{``Normal'' starbursts (so-called WR galaxies)}
Studies on WR galaxies (mostly BCD, Irr, spirals) are summarised in the 
reviews of Schaerer (1999ab). 
Including the detections of spectral signatures from both WN and WC stars
in a fair number of objects covering a large metallicity range, the following 
overall conclusions emerge from the studies of Schaerer \etal\ (1999a) and
Guseva \etal\ (2000). Except possibly at the lowest metallicities
a good agreement is found between the observations and the evolutionary synthesis
models of Schaerer \& Vacca (1998). From this comparison one finds clear
indications for short bursts ($\Delta t \le$ 2-4 Myr) in objects
with subsolar metallicity, an IMF compatible with Salpeter, and a 
large upper mass cut-off of the IMF, in agreement with several earlier 
studies. 
In addition, the observed WC/WN star ratios provide new constraints 
for mass loss and mixing scenarios in stellar evolution models
(Schaerer \etal\ 1999a).

We have recently undertaken a first study of metal-rich starbursts
(metallicities up to $\sim$ 2--3 times solar) with the aim of 
constraining the upper end of the IMF in such environments.
From the strengths of the observed WR features we derive a conservative
{\em lower limit} of \mup\ $\ga$ 30--40 \msun (Schaerer \etal\ 2000).
New observations are being obtained to improve the accuracy of
this result. 
Direct studies of the stellar content are of prime importance, also
to verify the reliability of indirect studies based on nebular line
analysis (cf.\ below). ***refs***
 
\subsection{Population synthesis studies using metallic lines}
Early starbursts studies using metallic lines have mostly concentrated
on the Calcium triplet (see references above).
More complete analysis of starbursts and AGN spectra using numerous 
optical stellar features have recently been presented by two groups.
Both approaches are based on ``classical'' population synthesis
using either observed stellar templates (Serote Roos \etal\ 1998, Boisson
\etal\ 2000) or cluster templates (Raimann \etal\ 2000ab).

The former authors analyse 5000 -- 8000 \AA\ spectra of 12 starburst,
Seyfert and LINERs, and use a synthesis technique yielding a
mathematically unique solution to determine the relative contributions
of different stellar populations. Regarding e.g.\ the importance of
super metal-rich stars their results differ from other work (e.g.\
Cid Fernandes \etal\ 1998, Gonz\'alez Delgado \etal\ 2000).

Raimann \etal\ have analysed average spectra of \hii\ galaxies,
starbursts and Sey2 taken from the Terlevich \etal\ (1991) catalogue.
Their main result regarding \hii\ galaxies is the finding of 
significant old ($\le$ 500 Myr) underlying populations, which
also modifies emission line diagnostics based on equivalent widths.
Their conclusions are supported by the study of selected BCD's 
by Mas-Hesse \& Kunth (1999) and by comparisons of predicted
and observed EL trends in several large samples of \hii\ galaxies
(Stasi\'nska \etal\ 2000). 

%%%%%%%%%%%%%%%%%%%%%%%%%%%%%%%%%%%%%%%%%%%%%%%%%%%%%%%%%%%%%%%
\section{Stellar features in the near--IR}
This spectral range is in most cases dominated by features from
late type stars (G, K, M) corresponding to red supergiants (RSG),
asymptotic giant branch stars (AGB), or red giants. Typical ages
for the appearance of these stars are $\ga$ 10$^7$, 10$^8$, and 10$^9$
yr respectively. 
The strongest features in the K and K band are
\begin{itemize}
\item atomic transitions
  of \atom{Si}{i} 1.59 $\mu$m,  \atom{Na}{i} 2.21,
 \atom{Fe}{i} 2.23, 2.24,  \atom{Ca}{i} 2.26,  \atom{Mg}{i} 2.28
\item molecular features, such as CO (6,3) 1.6, CO (2,0) 2.29
  and many OH lines (all wavelengths given in $\mu$m here).
\end{itemize}
The following papers (incomplete selection) describe their 
dependence on stellar type and luminosity class and/or provide 
spectral libraries:
Kleinmann \& Hall (1986), Lancon \& Rocca-Volmerange (1992), Origlia 
\etal\ (1993), Ali \etal\ (1995), F\"orster-Schreiber (1998).
Recently, some authors (e.g\ Gilbert \& Graham, these proceedings) 
have begun to use also theoretical spectra of cool stars for 
population studies.

Rather than providing a detailed review of the many studies
undertaken in this area, I will briefly recall some difficulties
affecting near-IR studies of stellar populations.
While obviously the traditional method of population synthesis,
decomposing the integrated spectrum in various template
constituents, can be equally applied to any wavelength range,
the near-IR properties predicted by evolutionary synthesis models 
are unfavourably affected by uncertainties in post main-sequence
stellar evolution and the modeling of cool stars.

Indeed, it is found that all current non-rotating stellar evolution 
models predict an incorrect variation of the relative red/blue 
supergiant lifetimes with metallicity $Z$ (Langer \& Maeder 1995)
\footnote{Depending on the adopted set, the predicted RSG/BSG
may well be correct for a certain metallicity (e.g.\ Geneva models
ok for solar metallicity). The predicted metallicity variation
does, however, not follow the observed trends.}. In addition
the predicted \teff\ of RSG may be too high (Origlia \etal\ 1999).
This implies, e.g.\ that the CO features predicted by evolutionary
synthesis models based on these tracks are weaker than observed
at subsolar $Z$ (Origlia \etal\ 1999).
Also, the predicted strong metallicity dependence of colors like V-K
(see Cervi\~no \& Mas-Hesse 1994) is therefore incorrect.
Improvements are expected from new stellar models allowing
for more realistic mixing scenarios including rotation 
(cf.\ Maeder \& Meynet 2000) and better understanding 
of mass loss in these phases.

Uncertain mass loss scenarii render the prediction of AGB stars, 
whose influence at long wavelength is non negligible for 
populations with ages 
$\ga$ 1 Gyr (e.g.\ Bruzual \& Charlot 1993), rather difficult, as 
best illustrated by Girardi \& Bertelli (1998).
For regions with small masses 
%$M \la$ 5--10 $\times 10^5$ \msun, 
stochastical effects (Lancon \& Mouhcine 2000) and fluctuations of the 
IMF from the finite number of stars (Cervi\~no \etal\ 2000) 
also lead to an expected dispersion. While these effects are
obviously of general nature, they are of particular importance
for predictions involving the dominant contribution from
stars with very short lived phases (e.g.\ AGB, WR).

%%%%%%%%%%%%%%%%%%%%%%%%%%%%%%%%%%%%%%%%%%%%%%%%%%%%%%%%%%%%%%%
\section{Combined stellar and nebular analysis of starbursts}
In various situations (e.g.\ starbursts with strong optical
EL; IR observations -- $\lambda \ga$ 4--10 $\mu$m where 
the continuum emission is dominated by dust and stellar
features are thus completely absent)
analysis of nebular emission lines are of interest to 
constrain the stellar population.

However, given the very nature of nebular physics, the EL are
not only sensitive to the ionising spectrum carrying information
of the stellar populations, but depend also strongly on the nebular
geometry and chemical composition. The dependence on these additional 
parameters (essentially the so-called ``ionisation parameter'' $U$ 
and composition)
render EL studies of stellar populations more complex
and require thus the use of sufficient observational constraints.

Recent developments and the current state-of-the-art of stellar
ionising fluxes forming the input to photoionisation models
have been reviewed by Schaerer (2000) and shall not be repeated here.
In the following we briefly summarise the main recent studies 
undertaken in the optical and IR (cf.\ Schaerer 2000 for more details.)

% % % % % % % % % % % % % % % % % % % % % 
\subsection{Optical studies}
Recent tailored starburst and photoionisation models are presented in 
the studies of Grac\ii a-Vargas \etal\ (1997: NGC 7714), Luridiana \etal\
(1999: NGC 2363, 2000: NGC 5461), Stasi\'nska \& Schaerer (1999: I Zw 18),
and Gon\'alez Delgado \& P\'erez (2000: NGC 604).
Although somewhat different in each study, the general approach is 
summarised in Grac\ii a-Vargas \etal\ (1997).

Overall one finds that both the stellar and nebular lines give
consistent results regarding the main properties of the stellar
population, such as age, IMF etc.
At a more detailed level, however, several of these studies encounter
significant difficulties (e.g.\ the temperature sensitive ratio 
[O~{\sc iii}] $\lambda$4363/5007 is underpredicted), which indicates 
that some physical are missing in the photoionisation models
(Stasi\'nska \& Schaerer 1999,  Luridiana \etal\ 1999, also
Stasi\'nska \etal\ 2000).
In short, 
although most observables can be reproduced by the combined starburst 
and photoionisation models --- and the tool can thus be used to derive
SB properties from the EL --- one has to conclude that for accurate
studies relying on nebular lines from \hii\ regions (and presumably also
more complex objects) some additional physical process(es) (possibly
shocks, conductive heating at X-ray interfaces etc.) must be taken 
into account (cf.\ Schaerer 2000).

% % % % % % % % % % % % % % % % % % % % % 
\subsection{Starburst + photoionisation models in the IR domain}
Analysis of IR observations (mostly from SWS and LWS on ISO) of starbursts 
based on combined SB + photoionisation models are just beginning to appear 
in the literature. In this context it is useful to keep some
intrinsic difficulties in mind are.
Given the nature of objects and the large apertures involved, the 
integrated spectrum generally includes a large variety of regions.
This fact, together with the complex geometries involved, render {\em a 
priori} the construction of photoionisation models difficult.

Simple models were constructed for case studies of Arp 299 and M82
by Satyapal \etal\ (1998) and Colbert \etal\ (1999) to interpret 
their LWS (40-200 \micron) spectra. Colbert \etal\ (1999) find
that the observed EL spectrum of M82 is compatible with an instantaneous
burst at ages $\sim$ 3--5 Myr, a Salpeter IMF, and a high upper mass
cut-off.
%value of $M_{\rm up} \sim$ 100 \msun.
Surprisingly, inspection of models with similar ingredients (cf.\
Stasi\'nska \& Leitherer 1996), show that the shorter wavelength data 
(see Genzel \etal\ 1998) is clearly incompatible with
the Colbert \etal\ model predicting too hard a spectrum. In view
of the few line ratios originating from the \hii\ gas and the large
number of free parameters the photoionisation model is underconstrained.
A larger wavelength coverage or other constraints are required.

F\"orster-Schreiber (1998) has described the geometry of clusters and
gas clouds in M82 by a single ``effective'' ionisation parameter.
% $U_{\rm eff}$.
This value has been adopted as typical for a sample of 27 starbursts in 
the SB + photoionisation models of Thornley \etal\ (2000). Instead of
modeling a simple stellar population their models are based on an ensemble 
of \hii\ regions following an observed luminosity function, which overall
leads to a reduction, albeit small, of the hardness of the ionising 
spectrum. From the ISO/SWS  [Ne~{\sc iii}]/[Ne~{\sc ii}] line
ratios they conclude that the observations are compatible with a high
upper mass cut-off ($M_{\rm up} \sim$ 50--100 \msun). To reproduce 
the relatively low average  [Ne~{\sc iii}]/[Ne~{\sc ii}] ratio, short
timescales of SF are required. 
More detailed studies including additional
observational constraints would be very useful to confirm this result.

A different approach has been taken by Schaerer \& Stasi\'nska (1999), who
modeled two well studied objects (NGC 5253, II Zw 40) with a fairly well
known massive star population and existing UV-optical-IR observations.
While their model successfully reproduces the stellar features and the 
observed ionisation structure of H, He, and O (as revealed from the optical
and IR lines), the predicted IR fine structure line ratios of
[Ne~{\sc iii}]/[Ne~{\sc ii}],  [Ar~{\sc iii}]/[Ar~{\sc ii}], and 
[S~{\sc iv}]/[S~{\sc iii}] show too high an excitation.
The origin of this discrepancy (atomic data? separate emission components ? 
other?) is still unknown.
In any case this attempt to describe two relatively ``simple'' objects
illustrates the current limitations and shows that further progress
is needed for a proper understanding and use of the IR fine structure
lines as reliable diagnostics.
Improvement is expected from multi-wavelength analysis of simpler objects 
(e.g.\ Galactic and LMC \hii\ regions, PN) and other ongoing work. 
Such studies should be crucial to reliably extend the diagnostic tools
to the IR to fully exploit the enormous observational capabilities
provided by recent and upcoming facilities in probing the properties
of massive star formation from the local Universe to high reshift.
%%%%%%%%%%%%%%%%%%%%%%%%%%%%%%%%%%%%%%%%%%%%%%%%%%%%%%%%%%%%%%%

%\begin{figure}[b]
%\begin{center}
%\includegraphics[width=.3\textwidth]{figure.eps}
%\end{center}
%\caption[]{Example of an electronically included eps-figure}
%\label{eps1}
%\end{figure}
%%%%%%%%%%%%%%%%%%%%%%%%%%%%%%%%%%%%%%%%%%%%%%%%%%%%%%%%%%%%%%%
\bigskip
{\em Acknowledgements} I thank the organisers for this very interesting
and stimulating workshop and for financial support.
Part of this work is also supported by the INTAS grant 97-0033.

%%%%%%%%%%%%%%%%%%%%%%%%%%%%%%%%%%%%%%%%%%%%%%%%%%%%%%%%%%%%%%%

%INDEX%%%%%%%%%%%%%%%%%%%%%%%%%%%%%%%%%%%%%%%%%%%%%%%%%%%%%%%%%%%%%%%
% Please check with the editor of your book whether he plans to
% include a "mutual" subject index - if so, please code your entries
% in the standard syntax. For your own purposes you may print your
% "personal" index by using the following commands:
%
%\clearpage
%\addcontentsline{toc}{section}{Index}
%\flushbottom
%\printindex
%%%%%%%%%%%%%%%%%%%%%%%%%%%%%%%%%%%%%%%%%%%%%%%%%%%%%%%%%%%%%%%%%%%%%

\end{document}